\documentclass[prb,aps,twocolumn,superscriptaddress]{revtex4-2}
\usepackage{graphicx}
\usepackage{dcolumn}
\usepackage{bm}
\usepackage{lipsum}
\usepackage{ragged2e}
\usepackage{braket}
\usepackage{bbm}
\usepackage{lipsum}
\usepackage{color}
\usepackage{xcolor}
\usepackage{amsfonts}
\usepackage{amsmath}
\usepackage{upgreek}
\usepackage{amsmath,amssymb,latexsym}
\usepackage{wasysym,stmaryrd,marvosym}
\usepackage{mathrsfs}
\usepackage{dsfont}
\usepackage{float}

\usepackage[mathlines]{lineno}
\usepackage[colorinlistoftodos]{todonotes}
\usepackage[colorlinks=true, allcolors=blue]{hyperref}

\begin{document}
 
\title{Valley-driven  {\it Zitterbewegung} in Kekul\'e-distorted graphene}

\author{Alex Santacruz}
\affiliation{Facultad  de  Ciencias,  Universidad  Aut\'onoma  de  Baja  California, 22800  Ensenada,  Baja  California,  M\'exico.}
\author{Priscilla E. Iglesias}
\affiliation{Facultad  de  Ciencias,  Universidad  Aut\'onoma  de  Baja  California, 22800  Ensenada,  Baja  California,  M\'exico.}
\author{Ramon Carrillo-Bastos}
\email{ramoncarrillo@uabc.edu.mx}
\affiliation{Facultad  de  Ciencias,  Universidad  Aut\'onoma  de  Baja  California, 22800  Ensenada,  Baja  California,  M\'exico.}
\author{Francisco Mireles}
\affiliation{Departamento de F\'isica, Centro de Nanociencias y Nanotecnolog\'ia, Universidad Nacional Aut\'onoma de M\'exico, Apdo. Postal 14, 22800 Ensenada, Baja California, M\'exico}
\date{\today}

\begin{abstract}
Graphene deposited on top of a Copper(111) substrate may develop a Y-shaped Kekulé bond texture (Kekulé-Y), locking the momentum with the valley degree of freedom of its Dirac fermions. Consequently, the valley degeneracy of its band structure is broken, generating an energy dispersion with two nested Dirac cones with different Fermi velocities. This work investigates the dynamics of electronic wave packets in the Kekulé-Y superlattice. We show that, as a result of the valley-momentum coupling, a valley-driven oscillatory motion of the wave packets ({\it Zitterbewegung}) could appear, but with a minor frequency than the {\it Zitterbewegung} effect associated with pristine graphene. Furthermore, we justify the presence of these {\it Zitterbewegung} frequencies in terms of the Berry connection matrix and a discrete symmetry present in the system. These results make Kekulé-Y graphene a compelling candidate for experimental observation of {\it Zitterbewegung} phenomenon in a two-dimensional system.

\end{abstract}

\maketitle

\section{Introduction} 

Experimental studies\cite{eom2020direct, PRL-KekO} have shown that inducing a Kekul\'e bond pattern in graphene leads to interesting changes in its band structure, as the breaking of its chiral (valley) symmetry\cite{Mudry2007,gutierrez2016imaging}, giving rise to appealing electronic and transport properties. The Kekul\'e-O (Kek-O)  type of bond distortion consists of
a periodic modification of alternated weak and strong (hexagonal) carbon bonding pattern in graphene. It leads to the coupling of the two distinct Dirac valleys\cite{chamon2000solitons} of its electronic band structure, and in momentum space, can be seen as effectively folding the pristine graphene Brillouin zone towards its center at the ${\Gamma}$ point\cite{cheianov2009hidden}. 
It has been theorized that a periodic arrangement of alkali atoms adsorbed in graphene can mimic the Kekule structure\cite{farjam2009energy,cheianov2009hidden}.
Indeed, this was recently confirmed experimentally\cite{PRL-KekO} where it was further observed that the Kek-O bonding drives the opening of a small bandgap. The latter is due to the broken chiral symmetry, which within tight-binding models, turns out as a consequence of the intervalley mixing\cite{Gamayun2018} and is proportional to the hopping amplitude $\Delta_0$ of the textured bonds. Another experimental study\cite{gutierrez2016imaging} found that epitaxial graphene grown on a Cu(111) substrate can develop a novel Y-shaped Kekul\'e distortion (Kek-Y). Later, a unified theoretical model of both Kekul\'e distortions\cite{Gamayun2018} showed that Kek-Y  also couples the valley isospin to the momentum, leading to a dispersion relation consisting of two concentric Dirac cones with different Fermi velocities. Notably, the Dirac fermions retain their massless, pseudo-relativistic nature from pristine graphene even though chiral symmetry is broken.

The physical consequences of both types of Kekule distortions on graphene electronic and transport properties have been the subject of exhaustive exploration theoretically. Spanning from studies of the enhanced valley-dependent Klein tunneling in a rectangular potential barrier in Kek-Y graphene by Wang et al.\cite{Barrier-Kek}, the coherent manipulation of a valley switch due to the selective valley pseudo-Andreev reflection in a  Kek-Y graphene superlattice potential by Beenakker et al.\cite{Beenakker-ValleySwitch}, to studies on the nature of valley-dependent currents in Kek-O distorted superconducting heterojunctions\cite{Wang2020}; also, recent theoretical studies by Andrade et al.\cite{Elias_2019} found that a uniaxial strain-driven Kek-Y distortion\cite{formation-strain-kek} in graphene can separate the Dirac cones away from the ${\Gamma}$-point. This was recently confirmed experimentally by Eom and Koo\cite{eom2020direct}. Other interesting studies include a device proposal for the manipulation of the valley orientation of channeled electrons through valley-induced precession produced by the Kek-Y texture\cite{wu2020electric} in graphene, as well as a study of the resonant electronic transport in graphene nanoribbons with periodic Kekulé distortions\cite{Elias2020}. More recently, the optical response and transport properties of Kek-Y graphene have been characterized using the Kubo\cite{KuboKek,Saul2021} and Boltzmann approaches\cite{Mojarro2020} as well as the addition of second order terms in the Kek-Y Hamiltonian\cite{andrade2021electronic,stegmann2021generalized}.

Until now, however, the dynamical properties of wave packets in Kek-Y monolayer graphene have not yet been investigated, nor the Zitterbewegung effect.
Schr\"odinger\cite{schrodinger1930kraftefreie} predicted that the dynamics of the Dirac Hamiltonian develop a high frequency oscillatory (trembling) motion (\textit{Zitterbewegung}). This effect is owed to the coupling of particle and antiparticle states of the Dirac system, yielding states that are superpositions of the two. The trembling occurs because the initial state is not an eigenstate of the coupled system. The {\it Zitterbewegung}, however, is not strictly a relativistic effect and comprises a broad class of phenomena that couple the momentum of quasiparticles to its spin or its spin-like degrees of freedom\cite{GeneralTheory-PRB2010}. Indeed, earlier theoretical studies on {\it Zitterbewegung} in solids were performed in superconductor materials by Lurie and Cremer\cite{Lurie&Cremer} and in bulk semiconductors by Cannata et al.\cite{Cannata} 
It was not after the works of Schliemann et al.\cite{schliemann2005zitterbewegung} and Zawadski\cite{Zawadski-PRB2005} on the plausibility of the observation of the {\it Zitterbewegung} in two-dimensional semiconductor systems with spin-orbit coupling that the topic experienced mainstream relevance. 
In connection with graphene, the wave packet dynamics and the {\it Zitterbewegung} effect was analyzed by Rusin and Zawadski\cite{rusin2007transient}, and later by Maksimova et al.\cite{maksimova2008wave}, showing that the effect and the intensity of the oscillations depends on the polarization of the sublattice pseudospin relative to the direction of motion. Subsequent studies have resorted to the dynamics of electronic wave packets to study the {\it Zitterbewegung} of a number of two dimensional materials and systems, including Dirac\cite{VillavicsPRB} and Weyl\cite{wpd-weyl} semimetals, silicene\cite{wpd-silicene,wpd-silicene2}, phosphorene\cite{wpd-phosphorene}, borophene\cite{wpd-borophene}, dice lattices\cite{wpd-aT3}, topological insulators\cite{wpd-ti,wpd-ti1,wpd-ti2}, in ABC-stacked $n$-layer graphene\cite{wpd-ABC}, and even more recently, in Moiré excitons in MoS$_2$/WSe$2$ heterobilayers\cite{PeetersgroupPRL2021}. Interestingly, the first direct experimental observation of the {\it Zitterbewegung} effect was realized not in high energy physics, nor in solid state materials, but in the realm of trapped ions\cite{gerritsma2010quantum}, using Bose-Einstein condensates\cite{leblanc2013direct} and spin-orbit-coupled ultracold\cite{SOCBEC} atoms of $^{87}$Rb. Most recently it was realized through classical laser optics simulation of the one-dimensional Dirac equation for a free particle\cite{OpticalSimulationZitter}.

In this work, we examine the dynamics of electronic wave packets generated by the chiral symmetry breaking in Kek-Y graphene, and study the {\it Zitterbewegung} phenomena associated to the simultaneous presence of the pseudospin- and valley-momentum coupling in this Dirac system. 

The outline of the rest of paper is as follows. In Sec.\,II we present the Dirac Hamiltonian for Kekul\'e-Y distorted graphene, and its associated band structure. In Sec.\,III We study the dynamics of the position, velocity and of the generalized spin matrix within the Heisenberg picture as well as the {\it Zitterbewegung} effect. Next we present the numerical studies and analysis of the dynamics of Gaussian wave packets under different initial momentum, pseudospin, and isospin conditions, as well as the {\it Zitterbewegung} of the averaged trajectories of the wave packets (Sec.\,IV). Lastly, we present our conclusions in Sec.\,V.

\section{The Kek-Y Hamiltonian and its two-flavor chiral fermions}
\begin{figure}[ht]
    \centering
    \includegraphics[width =0.7\columnwidth]{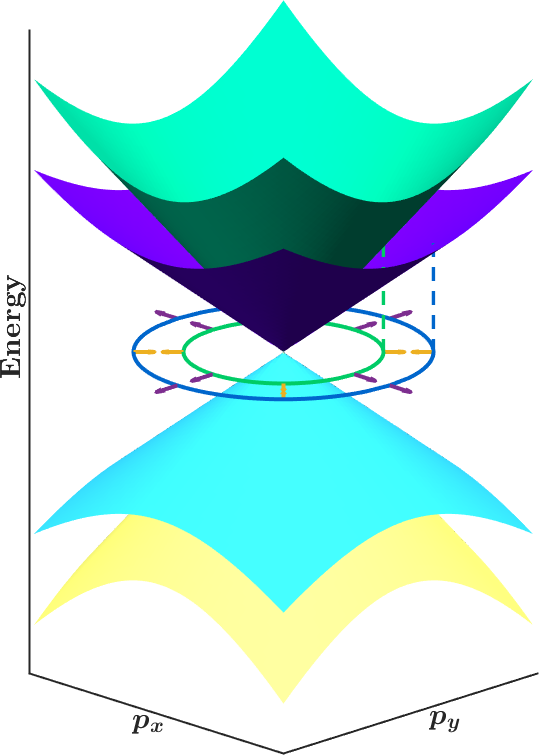}
    \caption{(Color online) Schematic dispersion relation for Kek-Y graphene described in Eq. (\ref{Eq:KekYDispersion}), and schematic depiction of the expected orientation of the sublattice pseudospin (purple arrows) and valley isospin (orange arrows) vectors for the positive energy (upper) Dirac cones. For the bottom cones, the direction of the pseudospin and isospin is reversed.}
    \label{fig:FiguraDispersionRelationAndSpinExpectationValues}
\end{figure}

Following Gamayun\cite{Gamayun2018}, we start by writing the Hamiltonian for the low-energy electronic excitations in graphene subject to Kekul\'e-Y textured bonding as
\begin{equation}\label{Eq:KekYHamiltonian}
    \hat H = v_\sigma \left(\boldsymbol{\hat S}^{\sigma}\cdot \boldsymbol{\hat p}\right) +  v_\tau \left(\boldsymbol{\hat S}^{\tau}\cdot \boldsymbol{\hat p}\right),
\end{equation}
where $\boldsymbol{\hat p}=(\hat p_x,\hat p_y,0)$ is the momentum vector, the sublattice ($v_\sigma$) and valley ($v_\tau$) Fermi velocities are renormalizations\cite{Gamayun2018} of the Fermi velocity ($v_f = 10~ \textrm{\AA}/\textrm{fs}$) of pristine graphene; without loss in generality, we take $v_\sigma = v_f$ and $v_\tau = \Delta_0 v_f$ , with $\Delta_0 < 1$ the Kekulé coupling strength parameter. We have used a set of generalized $4\times4$ pseudospin and isospin matrices to define the vector operators $\boldsymbol{\hat S}^\sigma = \tau_0\otimes\boldsymbol{\hat \sigma}$ and $\boldsymbol{\hat S}^\tau = \boldsymbol{\hat \tau}\otimes\sigma_0$, respectively; where the vectors $\boldsymbol{\hat \tau}=(\hat \tau_x,\hat \tau_y,\hat \tau_z)$ and $\boldsymbol{\hat \sigma} = (\hat \sigma_x,\hat \sigma_y,\hat \sigma_z)$ represent 3-component Pauli matrix vectors which act solely on the valley ($\tau$) and sublattice ($\sigma$) degrees of freedom, while the $\sigma_0$ and $\tau_0$ matrices are the corresponding $2\times2$ identity matrices. Note that, besides the pseudospin-momentum coupling term ($\boldsymbol{\hat S}^\sigma\cdot\boldsymbol{\hat p}$) akin to  pristine graphene, the Kek-Y distortion introduces an additional term ($\boldsymbol{\hat S}^\tau\cdot\boldsymbol{\hat p}$) that binds the valley isospin to the momentum (valley-momentum locking). The valley-locking term is responsible for breaking the valley degeneracy between the $K$ and $K'$ cones of pristine graphene, producing nested Dirac cones with different Fermi velocities, with the dispersion relation given by,
\begin{equation}\label{Eq:KekYDispersion}
   E_{\mu\nu}=\mu(v_{\sigma} +\nu v_{\tau})p,
\end{equation}
%
%
%
\noindent where $p = \sqrt{p_x^2 + p_y^2}$, with $\mu=+/-$ for the fermion/antifermion branch and $\nu=\pm$ are associated to the two distinct valleys. Although the valley-momentum coupling breaks chiral symmetry, the new chiral (valley) symmetry introduced by the Kek-Y superlattice\cite{Gamayun2018} still produces massless chiral fermions.
This superlattice chirality is actually a combination of two distinct flavors of chirality (or helicity), introduced by the pseudospin- and valley-momentum coupling. The sublattice helicity ($\frac{1}{2}\boldsymbol{\hat S}^{\sigma}\cdot \boldsymbol{\hat p}/p$) distinguishes positive energy states (fermions) whose pseudospin is parallel to the momentum from those of negative energy states (antifermions) whose pseudospin is antiparallel to the momentum. The valley chirality ($\frac{1}{2}\boldsymbol{\hat S}^{\tau}\cdot \boldsymbol{\hat p}/p$) differentiates between states whose isospin is parallel ($K$ cone) or antiparallel ($K'$ cone) to the momentum, which we shall call right and left chiral states, respectively. The chiral fermions in Kek-Y graphene can be distinguished by their associated group velocities. A right (left) chiral fermion (antifermion) will move at a group velocity $v_g = v_\sigma + v_\tau$ ($v_g = -v_\sigma - v_\tau$) while a left (right) chiral fermion (antifermion) will travel at group velocity $v_g = v_\sigma - v_\tau$ ($v_g = -v_\sigma + v_\tau$).  In Fig.\,\ref{fig:FiguraDispersionRelationAndSpinExpectationValues} we show a schematic draw of the resulting  dispersion laws and expected expected orientation of the sublattice pseudospin (purple arrows) and valley isospin (orange arrows) vectors at a given positive energy.
\\\\

\section{Heisenberg equations of motion and zitterbewegung}

To begin the analysis, we follow the standard procedure of calculating  the equations of motion of the quantum operators in the Heisenberg representation. The time evolution of an operator $\hat A$ with the Hamiltonian of
Eq.~(\ref{Eq:KekYHamiltonian}) is
\begin{equation}\label{Eq:KekYHeisenbergOperator}
    \frac{d}{dt}\hat A(t) = \frac{v_{\sigma}}{i\hbar}\left[\hat A(t),\boldsymbol{S}^{\sigma}\cdot \boldsymbol{\hat p}\right] + \frac{v_{\tau}}{i\hbar}\left[\hat A(t),\boldsymbol{S}^{\tau}\cdot \boldsymbol{\hat p}\right],
\end{equation}
where it is assumed that $\hat A$ does not depend explicitly on time $[\partial A/\partial t=0]$. Also, owing to chirality being a conserved property of massless fermions, we have omitted the time dependence of the momentum coupling terms. We can track the time evolution of the sublattice and valley degrees of freedom independently since $\left[S^\tau_i,S^\sigma_j\right]=0$ for all $i,j=x,y,z$. Let the index $\alpha$ represent either $\sigma$ or $\tau$, then the $4\times4$ matrix components of  $\boldsymbol{S}^\alpha$ satisfy
\begin{eqnarray}
    \left[S^{\alpha}_i,S^{\alpha}_j\right] & = & 2i\epsilon_{ijk}S^{\alpha}_k, \\
    \left[S^{\alpha}_i,p_{j}\right]~ & = & 0 ~~~~\text{for all}~i,j,k=x,y,z.
\end{eqnarray}
One can immediately notice that the Kek-Y phase conserves linear momentum ($\langle \dot{\boldsymbol{p}}\rangle = 0$). The time-dependence of the position operator $\boldsymbol{\hat r}=(\hat x,\hat y)$ is obtained by solving for the velocity operator, and it is given by
\begin{equation}\label{Eq:PositionDerivative}
    \boldsymbol{\hat v}(t)=\frac{d}{dt}\boldsymbol{\hat r}(t) = v_{\sigma}\boldsymbol{S}^{\sigma}(t) + v_{\tau}\boldsymbol{S}^{\tau}(t),
\end{equation}
from which it follows immediately that the total velocity operator is the sum of two velocity operators, one proportional to the time evolution of the sublattice and the other to the valley degree of freedom. Hence, the time-dependent quasiparticle velocities depend on the pseudospin and isospin, not just on its momentum.

By the same token, from the equation of motion (\ref{Eq:KekYHeisenbergOperator}) for a generalized spin matrix $\boldsymbol{S}^\alpha(t)$  we get
\begin{equation}
    \frac{d}{dt}\boldsymbol{S}^{\alpha}(t) = \frac{2v_{\alpha}}{\hbar} \boldsymbol{\hat p} \times \boldsymbol{S}^{\alpha}(t),
\end{equation}
which clearly describes the time-dependent pseudospin-momentum ($\alpha = \sigma$) and valley-momentum ($\alpha = \tau$) dynamical locking phenomena. This expression is equivalent to the equation of motion of a magnetic dipole in a magnetic field\cite{LessonsPRL} $\boldsymbol{B}$ (with $\frac{2v_f\boldsymbol{\hat p}}{\hbar}\rightarrow-\gamma\boldsymbol{B}$, being $\gamma$ the gyromagnetic ratio). The solutions for the components of $\boldsymbol{S}^{\alpha}(t)$, are given by
\begin{eqnarray}\label{Eq:SpinXTimeEvolution}
    S^{\alpha}_x(t) = && ~ S^{\alpha}_x
     + \frac{\hat p_y}{p}S^{\alpha}_z \sin\left(\omega_\alpha t \right) \nonumber \\
    &&+ \frac{\hat p_y}{p^2} \left( \hat p_x S^{\alpha}_y - \hat p_y S^{\alpha}_x \right) \left[ 1-\cos\left(\omega_\alpha t \right)\right], \\
    S^{\alpha}_y(t) = && ~ S^{\alpha}_y  \label{Eq:SpinYTimeEvolution}
     - \frac{\hat p_x}{p}S^{\alpha}_z \sin\left(\omega_\alpha t \right) \nonumber \\
    &&- \frac{\hat p_x}{p^2} \left( \hat p_x S^{\alpha}_y - \hat p_y S^{\alpha}_x \right) \left[ 1-\cos\left(\omega_\alpha t \right)\right],
\end{eqnarray}
here $S^{\alpha}_i$ (without the dependence of $t$) are the spin matrices in the Schr\"odinger picture. Eqs.\,(\ref{Eq:SpinXTimeEvolution}) and (\ref{Eq:SpinYTimeEvolution}) show that the pseudospin ($\alpha = \sigma$) and isospin ($\alpha = \tau$) vectors precess around the linear momentum with angular frequencies $\omega_\sigma = 2 p v_{\sigma}/\hbar$ and $\omega_\tau = 2 p v_{\tau}/\hbar$, respectively. The nature of these frequencies and their relation with the presence of a pseudospin and valley-driven {\it Zitterbewegung} effects will be discussed in more detail below. 

\begin{figure}[ht]
    \centering
    \includegraphics[width = 0.7\columnwidth]{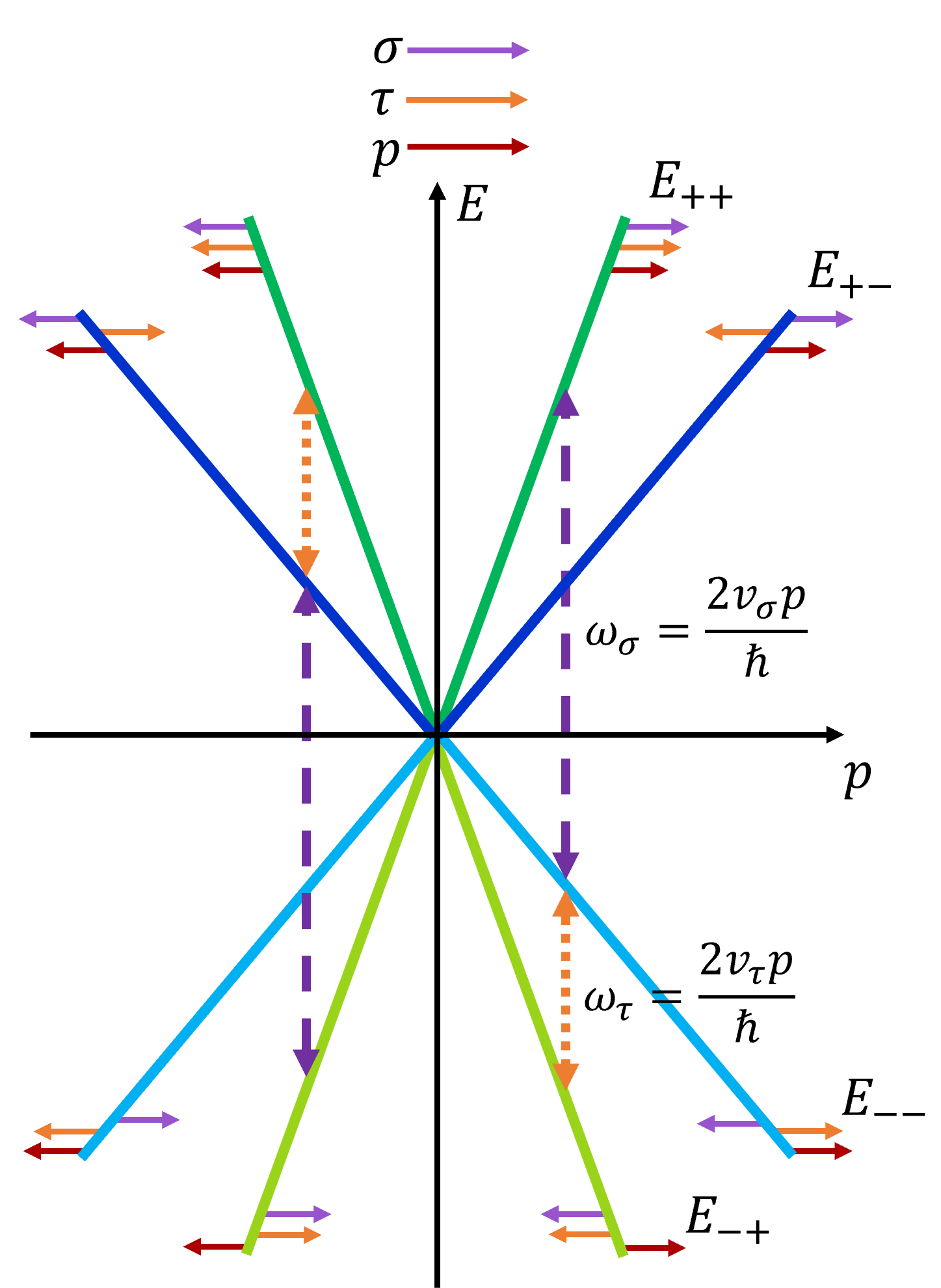}
    \caption{(Color online) Schematic cross section of the dispersion relation of the Kek-Y fermions, Eq.\,(1). The small horizontal arrows, indicates the direction of the momentum, and the expected value of the pseudo-spin and valley orientation. The oscillation frequencies of pseudospin- and valley-driven {\it Zitterbewegung} correspond to the energy difference between the cones connected by (purple) dashed and (orange) dotted double arrows, respectively.}
    \label{fig:FigureFrequenciesandDispersion}
\end{figure}

Once we substitute Eqs. (\ref{Eq:SpinXTimeEvolution}) and (\ref{Eq:SpinYTimeEvolution}) in (\ref{Eq:PositionDerivative}) and integrate over time accordingly, we obtain the time-dependent position vector operator. The explicit solution can be written as
\begin{equation}\label{Eq:PositionTimeEvolution}
    \boldsymbol{\hat r}(t) = \boldsymbol{\hat r}_0+\left(\boldsymbol{\hat v}_{0} + \boldsymbol{\hat v}^{\perp}\right) t + \boldsymbol{\hat \xi}(t)
\end{equation}
where $\boldsymbol{\hat r}_0 = \boldsymbol{\hat r}(0)$ is initial position vector operator, $\boldsymbol{\hat v}_0=\boldsymbol{\hat v}(0)$ is given by Eq. (\ref{Eq:PositionDerivative}), and the two time-independent components of $\boldsymbol{\hat v}^\perp$ are
\begin{equation}
    \hat v^\perp_x = \sum_{\alpha=\sigma,\tau} v_\alpha\frac{ \hat p_y}{p^2}\left(\hat p_x S^{\alpha}_y - \hat p_y S^{\alpha}_x \right),
\end{equation}
\begin{equation}
    \hat v^\perp_y = \sum_{\alpha=\sigma,\tau} v_\alpha \frac{\hat p_x}{p^2}\left(\hat p_x S^{\alpha}_y - \hat p_y S^{\alpha}_x \right).
\end{equation}
The second term in (\ref{Eq:PositionTimeEvolution}), $\left(\boldsymbol{\hat v}_{0} + \boldsymbol{\hat v}^{\perp}\right)t$, describes a rectilinear motion, akin to free particles, while the last term in Eq. (\ref{Eq:PositionTimeEvolution}), $\boldsymbol{\hat \xi}(t)=(\hat \xi_{x}(t),\hat \xi_{y}(t)),$ comprises the sum of two distinct oscillatory terms,
\begin{equation}
    \boldsymbol{\hat\xi}(t)=\boldsymbol{\hat \xi}^{\sigma}(t)+\boldsymbol{\hat \xi}^{\tau}(t),
\end{equation}
associated with the sublattice and valley degrees of freedom. These terms describe the trembling motion of a {\it Zitterwebegung} effect, where the components are determined by the expressions
\begin{eqnarray}
    \hat \xi^{\alpha}_x(t) = && \frac{\hbar}{2}\frac{\hat p_y}{p^2}S^{\alpha}_z \left[1 - \cos\left(\omega_\alpha t\right) \right]  \nonumber \\
    && - \frac{\hbar}{2}\frac{\hat p_y}{p^3}\left(\hat p_x S^{\alpha}_y - \hat p_y S^{\alpha}_x \right) \sin\left(\omega_\alpha t\right), \\
    \hat \xi^{\alpha}_y(t) = && -\frac{\hbar}{2}\frac{\hat p_x}{p^2}S^{\alpha}_z \left[1 - \cos\left(\omega_\alpha t\right) \right] \nonumber \\
    && + \frac{\hbar}{2}\frac{\hat p_x}{p^3}\left(\hat p_x S^{\alpha}_y - \hat p_y S^{\alpha}_x \right) \sin\left(\omega_\alpha t\right).
\end{eqnarray}

By evaluating the expectation value of the position operator in Eq.(\ref{Eq:PositionTimeEvolution}), it can be shown that a pseudospin-driven {\it Zitterwebegung} phenomenon arises due the mixing of states with same valley helicity, but opposite sublattice  helicity, and associated to the transition energies $E_{++}\leftrightarrow E_{--}$ and $E_{+-}\leftrightarrow E_{-+}$ 
whose absolute difference yields the oscillating frequency $\omega_{\sigma}(p)=2v_{\sigma}p/\hbar$.
This resembles the interference of positive and negative energy states that generate the Schr\"odinger {\it Zitterwebegung}\cite{maksimova2008wave} of the Dirac Hamiltonian.  In constrast, the valley-driven {\it Zitterwebegung} is owed to the mixing of states with opposite valley chirality (and same sublattice helicity), associated to the smaller transition energies $E_{++}\leftrightarrow E_{+-}$ and $E_{--}\leftrightarrow E_{-+}$ corresponding to the oscillating frequency $\omega_{\tau}(p)=2v_{\tau}p/\hbar$, 
as a direct result of the intervalley mixing introduced by the Kek-Y phase\cite{gutierrez2016imaging} (See Fig.\,2). Note that, unlike the former, the valley-driven {\it Zitterwebegung} occurs only between fermions or antifermions, that is, between states with the same energy sign.The valley-driven {\it Zitterwebegung} is thus more alike to that occurring in Rashba spin-orbit semiconductor two-dimensional systems\cite{schliemann2005zitterbewegung}. Note that other possible energy difference between $E_{+-}\leftrightarrow E_{--}$ and $E_{++}\leftrightarrow E_{-+}$ and that will correspond to the frequencies $2 p (v_{\sigma}-v_{\tau})/\hbar$ and $2 p (v_{\sigma}+v_{\tau})/\hbar$, respectively,  do not participate in the {\it Zitterbewegung} oscillations. The physical origin of the absence of these beating frequencies can be elucidated from the Berry phase connection matrix for the Kek-Y Hamiltonian system and its close relation with the {\it Zitterbewegung} amplitude oscillations in the position operator vector. Following D\'avid and Cserti\cite{GeneralTheory-PRB2010} on its general theory of the {\it Zitterbewegung}, such relation turns out evident by writing an equivalent form of the oscillatory terms of the position operator in terms of the Berry connection matrix,
\begin{equation}
 \boldsymbol{\hat\xi}(t)= \sum\limits_{k,l}(e^{i{\omega_{kl}}t}-1) { \boldsymbol A}_{kl}(\boldsymbol{p}) |\phi_k(\boldsymbol{p})\rangle\langle\phi_l(\boldsymbol{p})|,
\end{equation}
where $|\phi_k(\boldsymbol{p})\rangle$ are the eigenvectors of the Hamiltonian operator, 
$\omega_{k,l}=(E_k-E_l)/\hbar$ are the beating frequencies, and 
${\boldsymbol A}_{kl}(\boldsymbol{p})$ is the Berry connection matrix, defined as 
\begin{equation}
{\boldsymbol A}_{kl}(\boldsymbol{p}) =i\hbar \langle\phi_k(\boldsymbol{p})|\frac{\partial}{\partial \boldsymbol{p}} |\phi_l(\boldsymbol{p})\rangle.
\end{equation}
Note that nonzero value of the Berry connection matrix in Eq.(17) will entail in general a finite accumulated phase between the states involved, and as a consequence, from (16), a {\it multifrequency-Zitterbewegung} effect could arise in the system. 

Now, for the Kek-Y graphene Hamiltonian in Eq.(1) with eigenenergies $E_{1}=E_{++}$, $E_{2}=E_{+-}$, $E_{3}=E_{--}$ and $E_{4}=E_{-+}$, the frequencies defined before as $\omega_{\sigma}$ and $\omega_{\tau}$ will correspond precisely to $\omega_{13}$ and $\omega_{12}$, respectively. Using Eq.(17) in this system one can show that for the Kek-Y graphene Hamiltonian certain matrix elements vanishes; explicitly ${\boldsymbol A}_{14}(\boldsymbol{p})={\boldsymbol A}_{41}(\boldsymbol{p})=0$ as well as ${\boldsymbol A}_{32}(\boldsymbol{p})={\boldsymbol A}_{23}(\boldsymbol{p})=0$. This implies (from Eq.\,(16)) that the contribution to the {\it Zitterbewegung} associated to the beating frequencies $\omega_{14}$ and $\omega_{32}$ vanishes identically. 


This can be understood through symmetry arguments by considering the discrete symmetry operator $P=-\tau_x\otimes\sigma_x K$ (where $K$ is the complex conjugate) that commutes with the Kek-Y graphene Hamiltonian. This $P$-symmetry inverts the position operator ($ P\boldsymbol(i\hbar\frac{\partial}{\partial \boldsymbol{p}})P^{\dagger}= -i\hbar\frac{\partial}{\partial \boldsymbol{p}}$) and the \textit{Zitterbewegung} operator $\boldsymbol{\hat\xi}$. Therefore, it imposes a selection rule which makes the matrix elements between states with the same $P$-symmetry zero for both operators. Hence, no {\it Zitterbewegung} associated with the $E_{1}\leftrightarrow E_{4}$ and $E_{3}\leftrightarrow E_{2}$ energy differences is developed.


\section{Wave packet dynamics with valley-momentum coupling}

To further analyze the time-evolution of the position and velocity operators and the induced {\it Zitterwebegung} effects in Kek-Y graphene, we numerically study the dynamics of wave packets within the Schr\"odinger picture in real space. We consider an initial state ($t=0$) characterized by a four-component spinor with the form of a Gaussian wave packet given by
\begin{equation}\label{Eq:GaussianWavePacket}
    |\psi_{\sigma \tau}(\boldsymbol{x},0)\rangle = \frac{1}{d\sqrt{\pi}}e^{-\frac{|\boldsymbol{x}|^2}{2d^2} + i \boldsymbol{k_0}\cdot\boldsymbol{x}} |\tau\rangle \otimes |\sigma \rangle,
\end{equation} 
where $\boldsymbol{x} = (x,y)$ is the position vector, $d$ the width of the Gaussian wave packet. The initial wave vector $\boldsymbol{k}_0 = k_0(\cos\theta_p,\sin\theta_p)$ defines the average momentum vector ($\langle\boldsymbol{\hat p}_0\rangle = \hbar\boldsymbol{k_0} $) of the Gaussian wave packet, where $|\boldsymbol{k}_0|= k_0$ and $\theta_p$ is the polar angle of the momentum. The $|\tau\rangle$ and $|\sigma\rangle$ ket states describe the initial isospin and pseudospin polarization conditions, respectively. Using the Bloch sphere notation, they can be written in the basis of the eigenvectors of the $z$-Pauli matrix, $|z,\pm\rangle$ as
\begin{eqnarray}
     |\tau\rangle = &\,\, \cos\left(\theta_{\tau}/2\right) |z,+\rangle + e^{i\phi_\tau}\sin\left(\theta_{\tau}/2\right) |z,-\rangle, \\
    |\sigma\rangle = &\cos\left(\theta_{\sigma}/2\right) |z,+\rangle + e^{i\phi_\sigma}\sin\left(\theta_{\sigma}/2\right) |z,-\rangle
\end{eqnarray}
where the colatitudes and azimuthal angles of the isospin and pseudospin vectors are given by $\theta_\tau$, $\theta_\sigma$, $\phi_\tau$, and $\phi_\sigma$, respectively. 

We numerically evaluate the dynamics of the Gaussian wave packet (Eq.\ref{Eq:GaussianWavePacket}) by adopting the methodology in Ref.~\cite{strain2018}. In such numerical approach, a suitable, discretized time-evolution operator of the system is obtained using the Zassenhaus formula and Cayley expansion. This operator is used to compute the wave function at a later time $t+\Delta t$ at any point in space in terms of the wave function at a previous time $t$, starting from the initial condition given by Eq. (\ref{Eq:GaussianWavePacket}). In order to analyze the {\it Zitterbewegung}, we find convenient to use instead the Fourier transform of the Gaussian wave packet (Eq.\ref{Eq:GaussianWavePacket}) to compute the expectation values of the time-dependent velocity ($\boldsymbol{v}(t)$) and position ($\boldsymbol{r}(t)$) given by Eqs. (\ref{Eq:PositionDerivative}) and (\ref{Eq:PositionTimeEvolution}), respectively. Unless specified otherwise, in all the simulations shown here we considered a Kekul\'{e} coupling amplitude of $\Delta_0 = 0.1$.

\begin{figure}[h]
    \centering
    \includegraphics[width=\columnwidth]{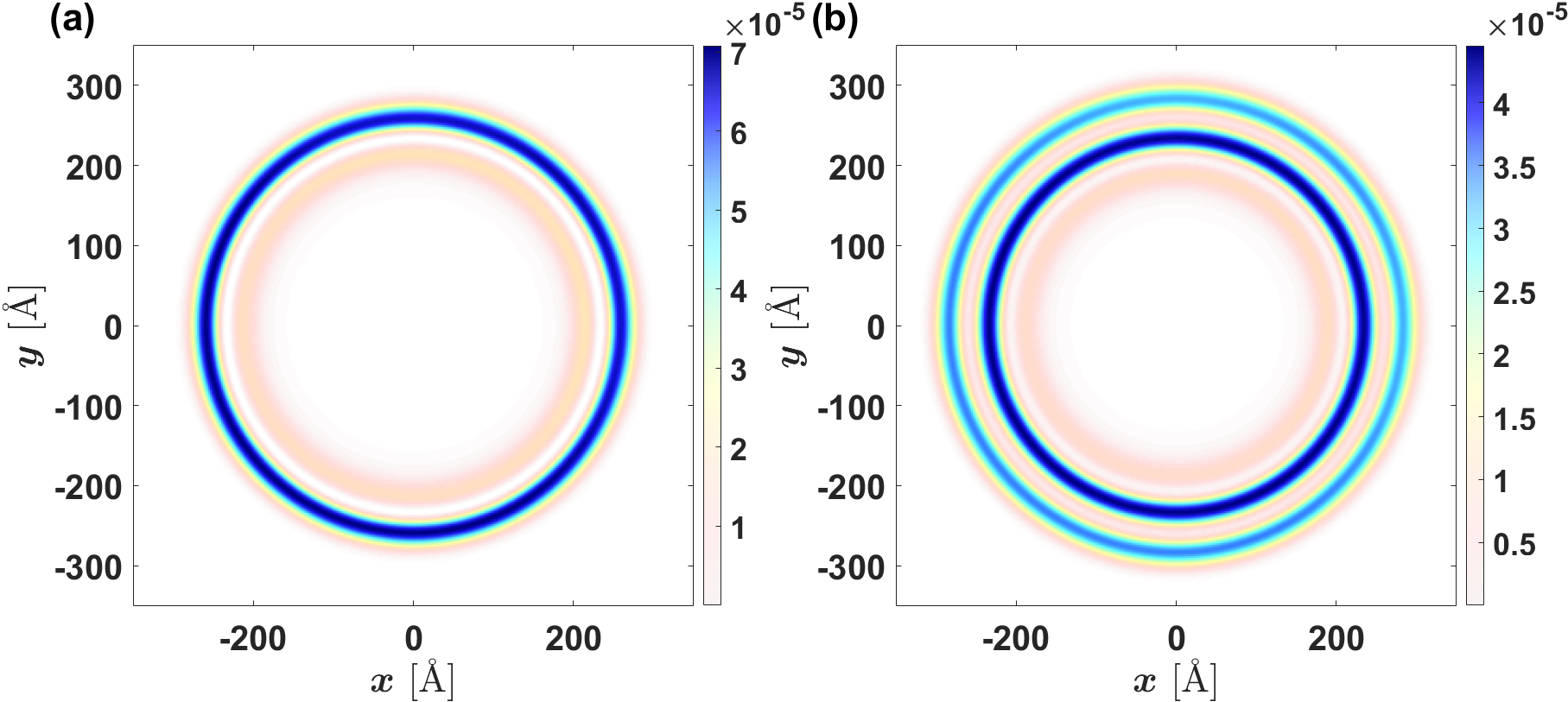}
    \caption{Snapshots of the probability density at $t=25~\text{fs}$ developed by a stationary ($k_0 = 0$)  Gaussian packet of initial width $d = 20~\text{\AA}$ in pristine (a) and Kek-Y (b) graphene ($\Delta_0 = 0.1$), with $\theta_\sigma = \theta_\tau = 0$. Note that Kek-Y graphene case develops two concentric ring of high probability density.}
    \label{fig:FigurePD_NoMomentum}
\end{figure}

First, for illustrative purposes of our quantum simulations, we compare the dispersion of a Gaussian wave packet of width $d = 20~ \text{\AA}$ at $t = 25~\text{fs}$ in pristine [Fig. \ref{fig:FigurePD_NoMomentum}(a)] and Kek-Y graphene [Fig. \ref{fig:FigurePD_NoMomentum}(b)] with zero average momentum ($\langle\boldsymbol{\hat p}(0)\rangle =0$) and both pseudospin and isospin vectors pointing in the positive $z$ direction, that is $\theta_\sigma = \theta_\tau = 0$. This initial condition is equivalent to a superposition of every possible state with momentum, pseudospin, and isospin states in the $xy$ plane.
As such, the wave packet contains left and right chiral fermionic and antifermionic states, which disperse uniformly in every direction due to pseudospin- and valley-momentum locking. As a result, the wave packet in pristine graphene shown in Fig. \ref{fig:FigurePD_NoMomentum}(a) forms a probability density ring that propagates isotropically outwards from its center at an associated group velocity close to the Fermi velocity $v_f$. In Kek-Y graphene [Fig. \ref{fig:FigurePD_NoMomentum}(b)], the valley degeneracy allows the same wave packet to travel at two different Fermi velocities; hence the wave packet develops two concentric probability density rings. We associate the outermost ring with right chiral fermions and antifermions that travel at group velocity $v_g = \pm (v_\sigma + v_\tau)$, and the slower ring with left chiral fermions and antifermions that propagate with group velocity $v_g = \pm (v_\sigma - v_\tau)$. From quantum simulations of the time evolution of Gaussian wave packets in a Dirac system, it is well known the appearance of the rippling behavior of the probability density \cite{thaller2004visualizing} of the wave packet as time passes while maintaining zero average position. 
In Kek-Y graphene, the rippling effect forms a (light pink) ring shadow that travels behind the main probability density rings.


\begin{figure}[h]
    \centering
    \includegraphics[width=0.8\columnwidth]{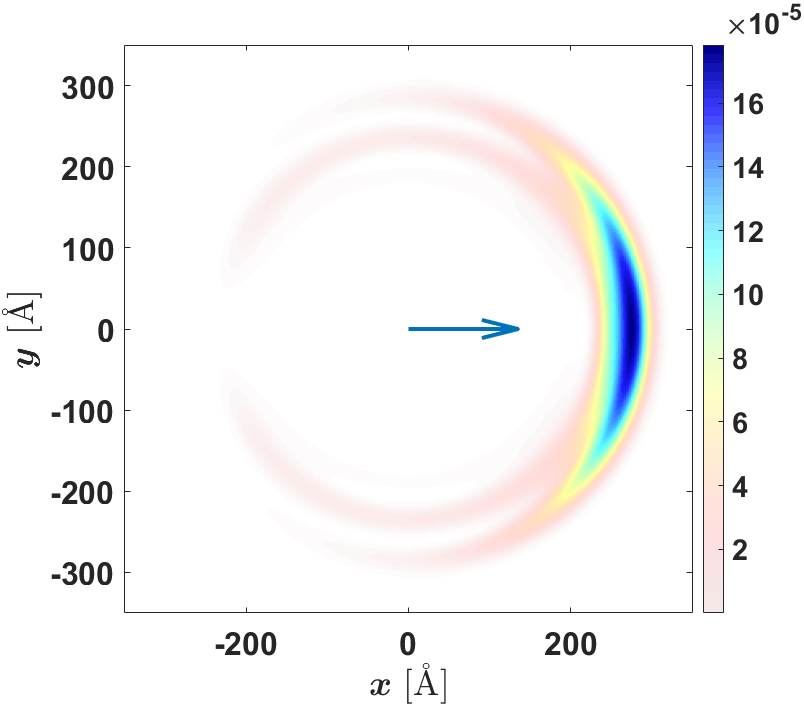}
    \caption{Probability density of a Gaussian wave packet of width $d=20~\textrm{\AA}$ at $t=25~\text{fs}$ in Kek-Y graphene ($\Delta_0=0.1$) with average momentum ($k_0 = 0.05~\text{\AA}^{-1}$, $\theta_p=0$)  parallel to the isospin and pseudospin in the $x$ direction, indicated by the blue arrow, ($\theta_\tau = \theta_\sigma = \pi/2$, $\phi_\tau = \phi_\sigma = 0$).
    \label{fig:FigureDynamicEigenstateV3}}
\end{figure}

Now, in Fig. \ref{fig:FigureDynamicEigenstateV3}, we consider a Gaussian wave packet of width $d=20~\textrm{\AA}$ in Kek-Y graphene whose momentum ($k_0 = 0.05~\textrm{\AA}^{-1}$), and the pseudospin and isospin point in the same direction, specifically along the $x$ direction ($ \theta_p = 0 $, $ \theta_\sigma = \theta_\tau = \pi/2 $, and $ \phi_\sigma = \phi_\tau = 0 $). By fixing the average momentum of the wave packet parallel to the pseudospin and isospin vectors at $ t=0~\text{fs}$, the time-evolved wave packet propagates as a whole in the direction of the momentum (represented by a blue arrow). From the plot one can estimate the average position of the wave packet at this time ($ t=25~\text{fs} $) that leads us to estimate that the wave packet is moving at an average velocity, very close indeed to the group velocity of the right chiral fermions ($\approx v_\sigma + v_\tau$). Due to the presence of states with momentum orthogonal to the pseudospin and isospin, a pair of crescent-shaped trails form behind the Gaussian and the average velocity fluctuates rapidly with time in the direction of motion.

\begin{figure}[h]
    \centering
    \includegraphics[width=\columnwidth]{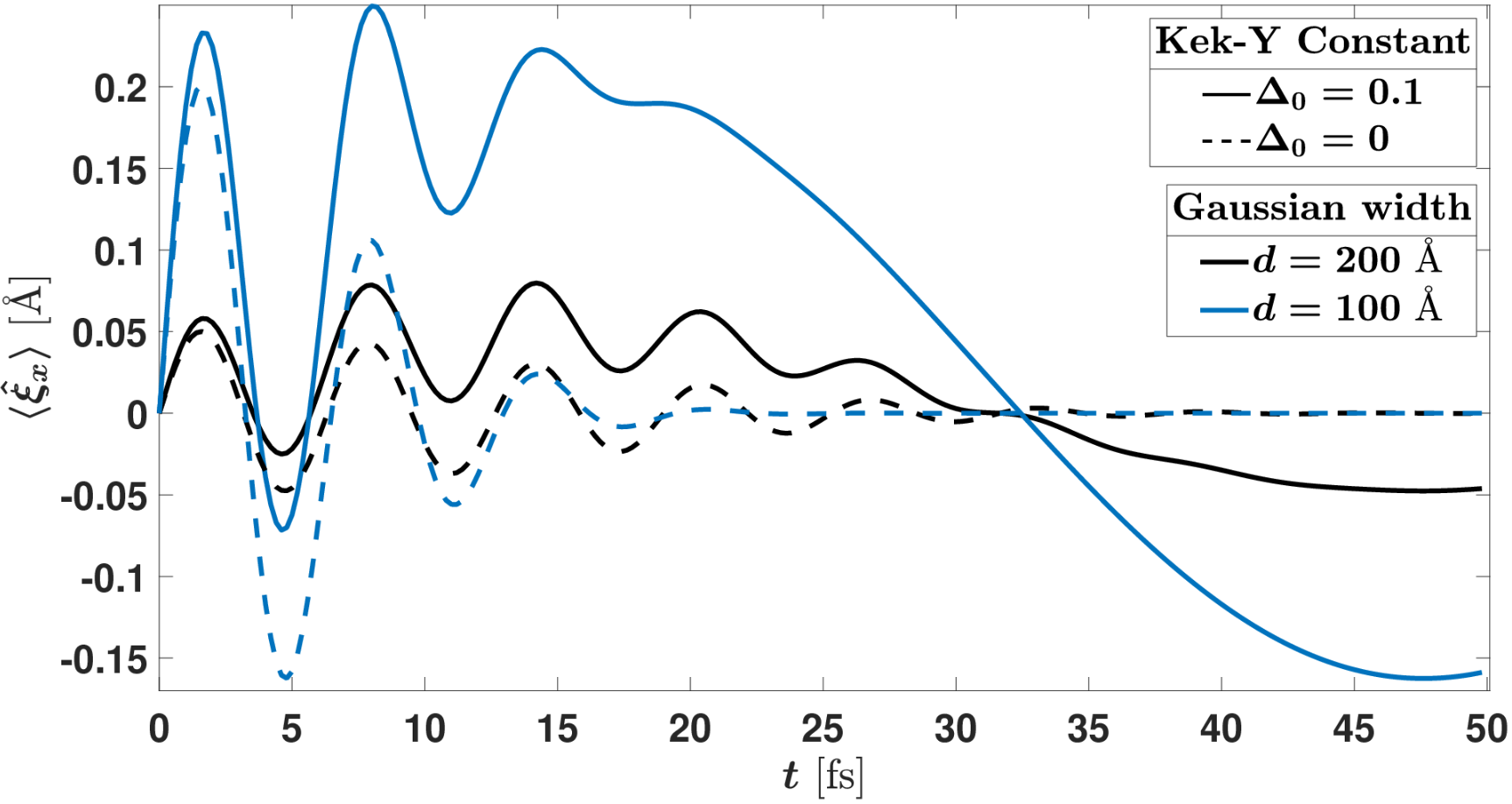}
    \caption{Numerical evaluation of the oscillatory part for the expectation value of the position of the centroid of the wave packets as a function of time for Kek-Y (solid curves) and pristine (dashed curves) graphene. The initial Gaussian wave packet has an average momentum ($k_0 = 0.05~\text{\AA}^{-1}$, $\theta_p=0$) parallel to the isospin and pseudospin in the $x$ direction ($\theta_\tau = \theta_\sigma = \pi/2$, $\phi_\tau = \phi_\sigma = 0$). Two cases are considered, Gausssian widths of  $d = 200~\textrm{\AA}$ (black) and $d = 100~\textrm{\AA}$ (blue). Two types of longitudinal {\it Zitterbewegung}  naturally emerge when the Kek-Y distortion is present. The oscillations with the smaller period are due the pseudospin induced {\it Zitterbewegung}, whiles the larger superimposed period is due the valley-driven {\it Zitterbewegung}. }
    \label{fig:FigureZitterEigenstate}
\end{figure}

In Fig. \ref{fig:FigureZitterEigenstate}, we characterize the resulting trembling motion of the Gaussian wave packets by plotting only the oscillatory part for the expectation value of the position of the centroid of the wave packets. For the simulations, we consider two different widths of the initial wave packets ($d=100~\textrm{\AA}$ and $d=200~\textrm{\AA}$) for both Kek-Y graphene (solid curves) and pristine graphene (dashed curves), and set the average momentum ($k_0 = 0.05~\text{\AA}^{-1}$, $\theta_p=0$), parallel to the initial pseudospin and isospin vectors ($ \theta_p = 0 $, $ \theta_\sigma = \theta_\tau = \pi/2 $, and $ \phi_\sigma = \phi_\tau = 0 $), that is along the $x$-axis. Note that for this case there is no {\it Zitterbewegung} in the $y$ direction, that is $\langle \hat \xi_y(t) \rangle = 0$. However, the components of the wave packet whose momentum is orthogonal to pseudospin and isospin induce a so-called \textit{longitudinal} {\it Zitterbewegung}\cite{maksimova2008wave} in the direction of propagation, as expected from the previous theoretical analysis, see Eqs. (14) and (15). As discussed, in pristine graphene, the interference between states of opposite sublattice helicity with respective positive and negative energies drives the pseudospin {\it Zitterbewegung} with characteristic frequency $\omega_\sigma = 2 v_\sigma p/\hbar$; while in  Kek-Y graphene, the interference between states with opposite valley chirality corresponding to positive and negative energies induces an additional oscillatory motion, but in this case with a much lower frequency $\omega_\tau = 2 v_\tau p/\hbar$, that is $\omega_\tau = \Delta_0 \omega_\sigma$. We can estimate the expected characteristic frequencies by using $p\rightarrow \hbar k_0$, with $k_0= 0.05~\text{\AA}^{-1}$, which results in characteristic half-period oscillations of exactly $\pi$\,fs and $10\pi$\,fs, respectively, (with  $T_{\sigma} = 2\pi/\omega_\sigma$ and $T_{\tau} = 2\pi/\omega_\tau$). As the wave packet dynamics shows in Fig. \ref{fig:FigureZitterEigenstate}, in the absence of Kek-Y distortion ($\Delta_0=0$) an oscillatory behavior is developed of $\langle \hat \xi_x(t) \rangle$, from which we can extract a half-period of approximately 3 fs with a decaying behavior of its amplitude as time increases. However, when the Kek-Y distortion is present ($\Delta_0=0.1$) a second oscillatory behavior is superimposed to the first one with a half-period of approximately of 32 fs. It is clear that such values are in excellent agreement with the expected theoretical estimates for the characteristic pseudospin and valley \textit{longitudinal} {\it Zitterbewegung} oscillations periods discussed above.



\begin{figure*}[t]
    \centering
    \includegraphics[width=0.8\textwidth]{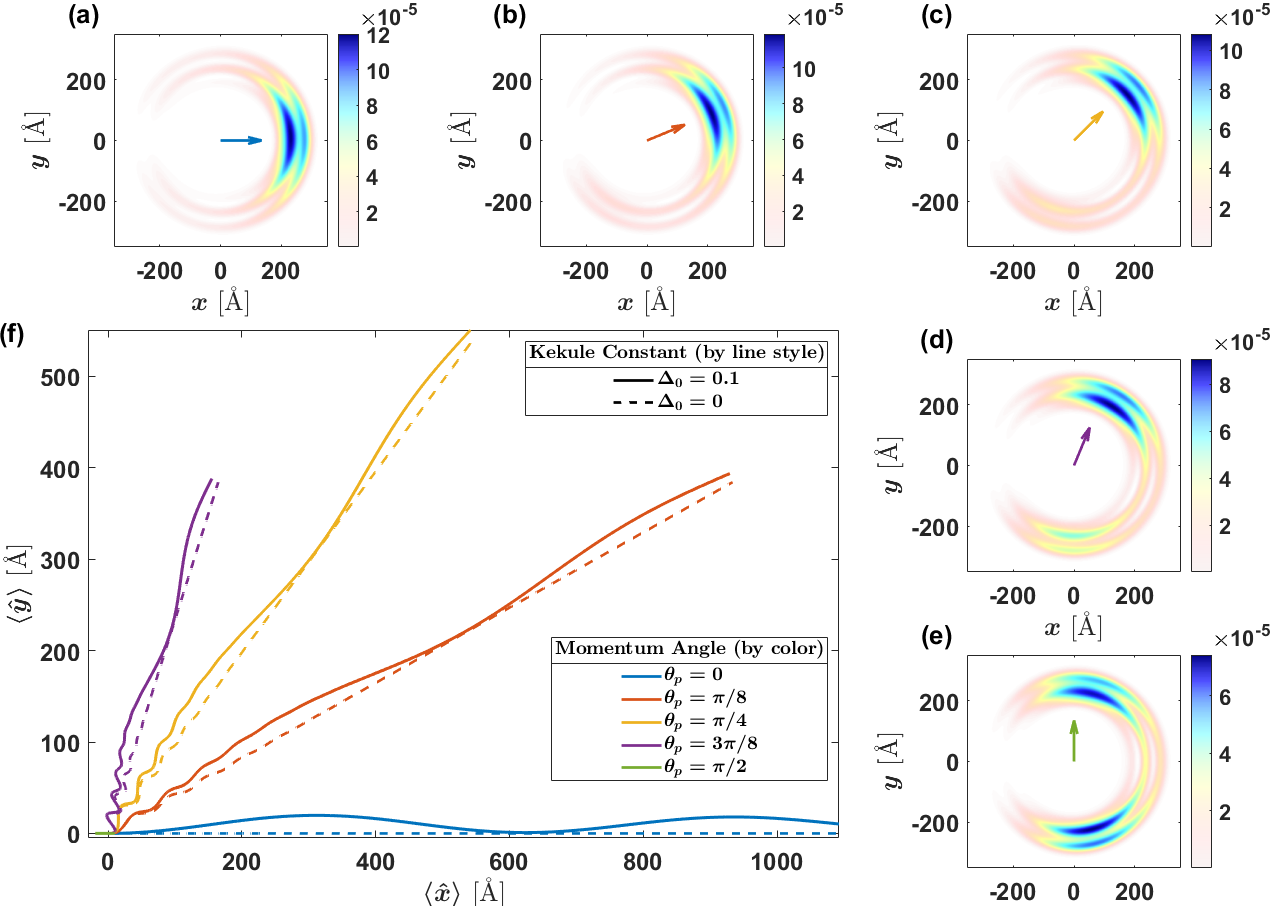}
    \caption{Dynamics of Gaussian wave packets with for various directions of momentum given by $\theta_p$ ($k_0 = 0.05~\textrm{\AA}^{-1}$) with the pseudospin and isospin, initially, in the positive $x$ and negative $z$ directions, respectively ($ \theta_\sigma = \pi/2, \phi_\sigma = 0 $, $\theta_\tau = \pi $). (a)-(e) Probability densities at $t=25~\text{fs}$ corresponding to Gaussian wave packets of width $d=20~\textrm{\AA}$, where the momentum vectors are represented by arrows whose color corresponds to the color of the curves in (f). (f) Average trajectories of the Gaussian wave packets of width $d=200~\textrm{\AA}$ in Kek-Y (solid lines) and pristine graphene (dashed lines). The final time of the integration is $t_f=110~\text{fs}$.}
    \label{fig:FigureDynamicsVariousMomentaV2}
\end{figure*}

The decaying nature of the amplitude at long times of the {\it Zitterbewegung} oscillations is well understood. Early studies by Rusin and Zawadzki \cite{rusin2007transient,rusin2008zitterbewegung}, focused on the dynamics of wave packets in mono- and bilayer graphene, show that {\it Zitterbewegung} phenomenon persists so long as the wave subpackets (associated to particle and anti-particle states) are physically overlapped in real space, decaying as the subpackets move away from each other. Thus it can be argued that the decaying in time of {\it Zitterbewegung} is a decreasing function of its spatial uncertainty. From Fig. \ref{fig:FigureZitterEigenstate}, we can infer that the decay time depends on the sublattice ($v_\sigma$) and valley ($v_\tau$) velocities, as the valley {\it Zitterbewegung} decays much slower than its pseudospin counterpart. This is clearly due to the fact that $v_\tau/v_\sigma = \Delta_0$. Since in this case, $\Delta_0 = 0.1$, therefore the valley {\it Zitterbewegung} decays ten times slower than pseudospin induced {\it Zitterbewegung}.

\begin{figure}[h]
    \centering
    \includegraphics[width = \columnwidth]{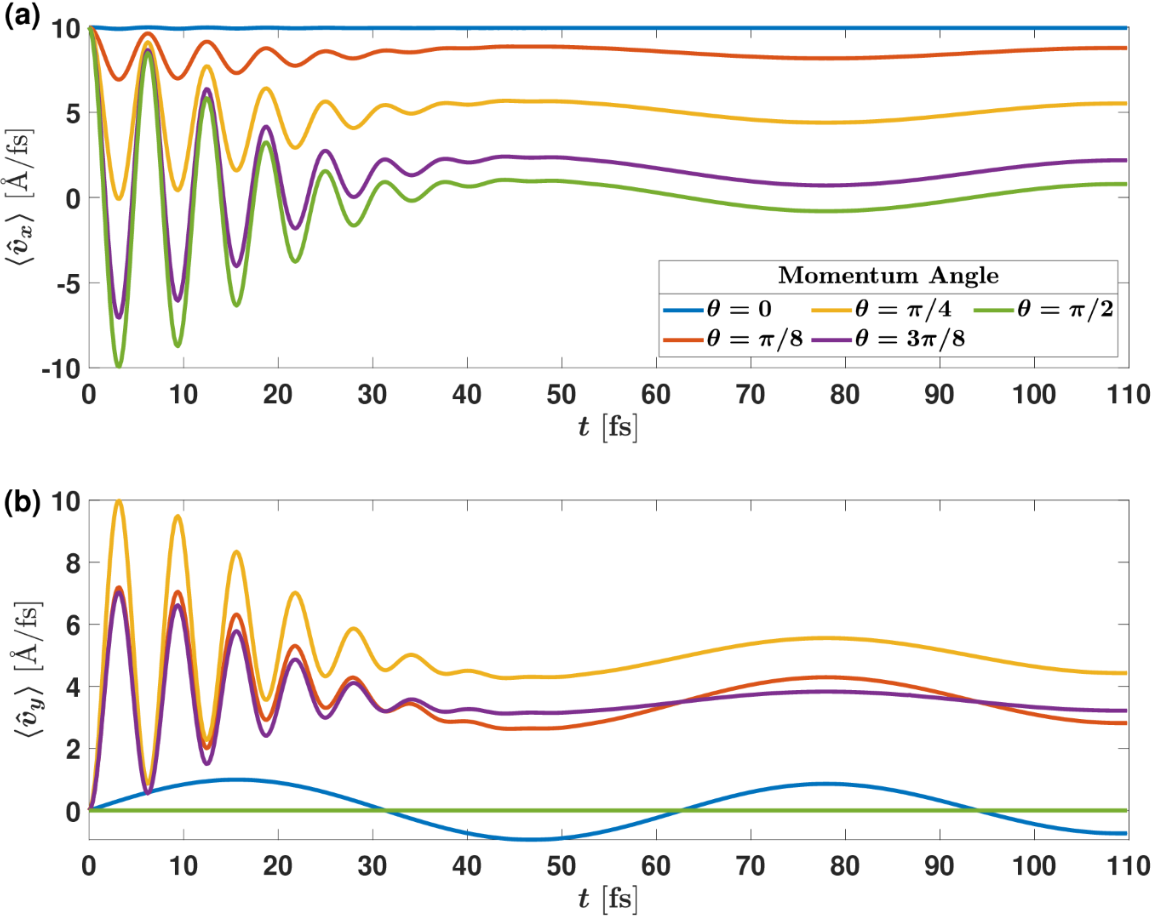}
    \includegraphics[width = \columnwidth]{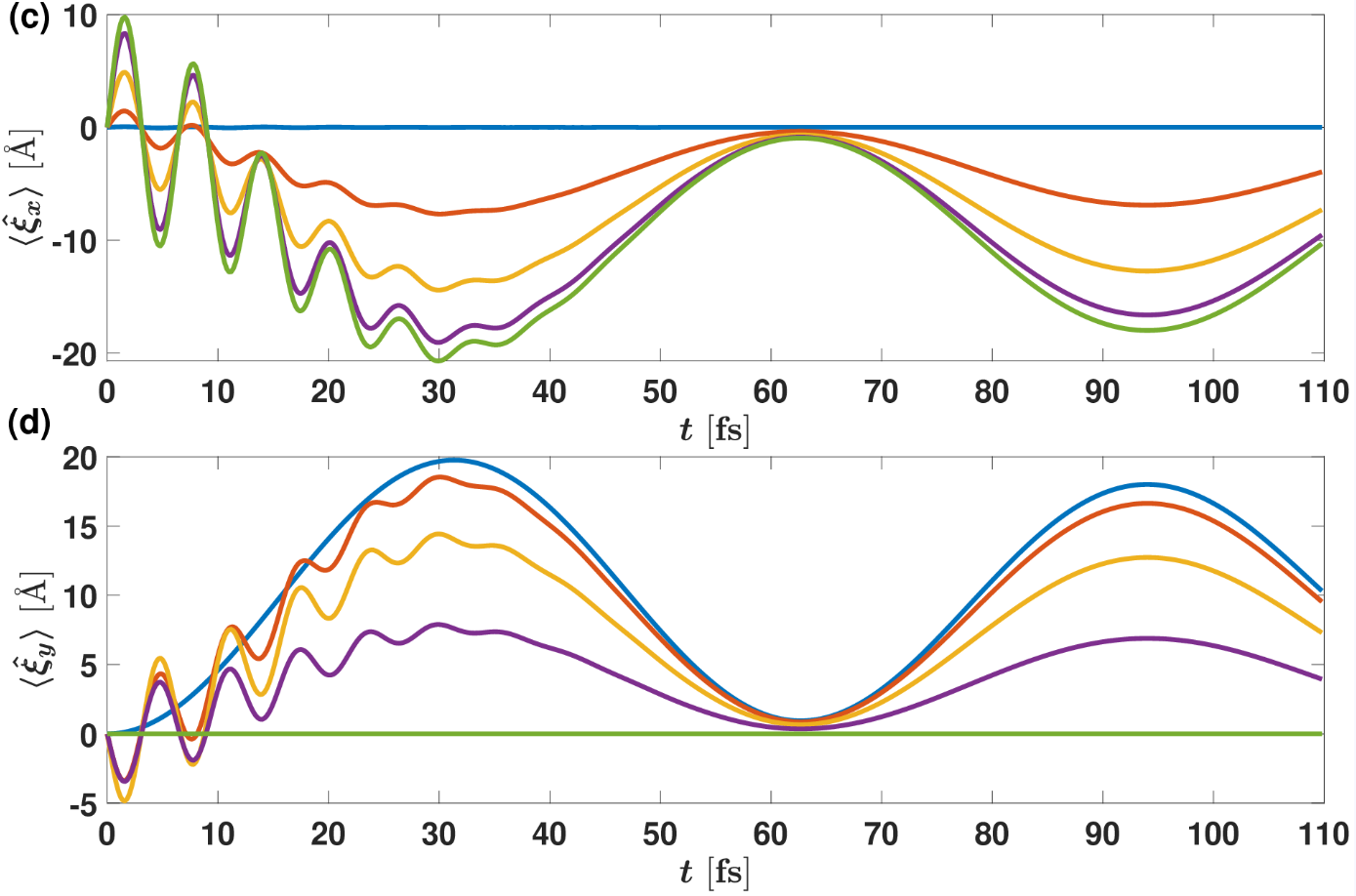}
    \caption{Time evolution of the average velocities (a) $\langle v_x(t)\rangle$ and (b) $\langle v_y(t)\rangle$, as well as the \textit{Zitterbewegung} in the (c) $x$ and (d) $y$ direction for Gaussian wave packets of width $d = 200~\text{\AA}$ in Kek-Y graphene for various directions of momentum, $\theta_p$. The pseudospin and isospin are initially in the positive $x$ and negative $z$ directions, respectively, that is $ \theta_\sigma = \pi/2, \phi_\sigma = 0 $, and $\theta_\tau = \pi $.}
    \label{fig:FiguraVelocityZitterVariosAngulosMomentoV2}
\end{figure}

In Fig. \ref{fig:FigureDynamicsVariousMomentaV2}, we show the time-evolved Gaussian wave packets of width $d=20~\textrm{\AA}$ in Kek-Y graphene at time $t=25~\text{fs}$ for various directions of momentum, $\theta_p$, while maintaining the pseudospin fixed in the $x$ direction ($\theta_\sigma = \pi/2, \phi_\sigma = 0$) and setting the isospin in the negative $z$ direction ($\theta_\tau = \pi$), with $k_0 = 0.05~\textrm{\AA}^{-1}$. The average momentum vector of the incident Gaussian wave packet is represented by arrows, where its  color corresponds to the values of $\theta_p$ described in Fig.\,\ref{fig:FiguraVelocityZitterVariosAngulosMomentoV2}. In Fig.\,\ref{fig:FigureDynamicsVariousMomentaV2}(a), the average momentum is aligned with the initial pseudospin vector ($\theta_p = 0$) and the wave packet propagates horizontally towards the right. Here, similarly as it occurs in Fig. \ref{fig:FigurePD_NoMomentum}(b), the probability density split into two wave fronts due to their different group velocities in Kek-Y graphene, but this time its cylindrical symmetry is lost and the wave packets develop maximums in the probability density which are drifted to the right due that its eigenstate solutions with positive velocities are dominant owing that the initial momentum $\langle\boldsymbol{p}_0\rangle$ of the wave packet is to the right. As we increase the angle $\theta_p$ in Fig. \ref{fig:FigureDynamicsVariousMomentaV2}(b)-(e), we observe that the centroid of the wave packets tends to follow the direction of motion set by $\langle\boldsymbol{p}_0\rangle$. However as the $\theta_p$ approaches to $\pi/2$, that is, when the direction of the momentum tends to be aligned to polarization of the pseudospin ($\theta_{\sigma}$), the wave packets splits into two groups of two maximums (associated to the valley and pseudospin components propagating in opposite directions. Hence, as $\theta_p$ increases, the probability density of the component wave packet with pseudospin antiparallel to momentum grows in amplitude, and moves in the opposite direction $\langle \boldsymbol{p}_0\rangle$.
The asymmetry of the probability density along the $x$ axis being maximum for positive $x$ show us that the wave packet is indeed propagating towards that direction as well. This is most apparent when the pseudospin is precisely orthogonal to the momentum, as in Fig. \ref{fig:FigureDynamicsVariousMomentaV2}(e). A similar behavior was described in Ref.\,[ \cite{maksimova2008wave}] for the dynamics of wave packets in monolayer graphene.

In Fig. \ref{fig:FigureDynamicsVariousMomentaV2}(f), we plot the corresponding trajectory described by the expectation values of the components of the position operator as time evolves ($\langle \hat x(t) \rangle ,\langle \hat y(t) \rangle $), in Kek-Y (solid lines), and in pristine graphene (dashed lines), and for different angles $\theta_p$ (the rest of the parameters are as in Fig.\,5(a)). Clearly a uniform rectilinear path along the $x$-axis are followed by the quasiparticle wave packets with an average velocity of 10 fs/\AA, which corresponds to the Fermi velocity of the quasiparticles in graphene without Kekule-distortion. Once the Kek-Y bond texture is introduced a longitudinal valley {\it Zitterbewegung} is induced.

For $\theta_p = 0$, the average rectilinear motion of the wave packet in Kek-Y graphene (blue solid line) is very close to the path of the same wave packet in pristine graphene (blue dashed line) as $ \langle \boldsymbol{v}_0 + \boldsymbol{v}^{\perp}\rangle \approx v_\sigma $. The $v_x(t)$ operator exhibits longitudinal pseudospin precession similar to Fig. \ref{fig:FigureZitterEigenstate}, however the amplitude of this oscillation is very small compared to the other cases. Meanwhile, the $v_y(t)$  component departs from zero and begins to oscillate with frequency $\omega_\tau$ as the isospin begins to precess since $|\tau\rangle = |z,-\rangle$ is perpendicular to $\langle \boldsymbol{p} \rangle$.

As $\theta_p$ increases, the average trajectories in Fig. \ref{fig:FigureDynamicsVariousMomentaV2}(f) roughly follow the direction of the momentum with the wave packet traveling a similar distance in both Kek-Y and pristine graphene since $v_\tau/v_\sigma = \Delta_0=0.1$. The amplitude of the precession in the $x$ direction given by $\langle v_x(t) \rangle$ increases and, as the the precession decays, the velocity vector $\langle \boldsymbol{v}(t)\rangle$ begins to settle at a constant value given by $\langle \boldsymbol{v}_0 + \boldsymbol{v}^{\perp} \rangle $. However, because $\langle\boldsymbol{v}_{0,y}\rangle = 0$, the magnitude of average velocity post decay $|\langle\boldsymbol{v}(t\rightarrow\infty)\rangle|$ decreases as $\theta_p$ increases. Therefore, the average radial distance traveled by the wave packet reduces from $\sim v_\sigma t$ for $\theta_p=0$ to zero for $\theta_p=\pi/4$. In the Schrodinger picture (Figs. \ref{fig:FigureDynamicsVariousMomentaV2}(a)-(e)), we can attribute this to the probability amplitude of states that propagate in the direction opposite to $\langle\boldsymbol{p}\rangle$ (those with opposite chirality and helicity) increasing as a function of $\theta_p$.

Now, we proceed to explore further the {\it Zitterbewegung} using the analytical expressions for the velocity and position operators derived in Sec.\,III by calculating its expectation values through direct numerical integration. In Figs.\,\ref{fig:FiguraVelocityZitterVariosAngulosMomentoV2}(a)-(b), we plot the expectation values of the velocity components as a function  of time for several directions of the momentum by varying $\theta_p$. The rest of the parameters used were as follows; $d = 200~\text{\AA}$, $k_0 = 0.05~\textrm{\AA}^{-1}$, $\Delta_0 =0.1$, while maintaining the pseudospin and isospin fixed in the $x$ direction ($\theta_\sigma = \pi/2, \phi_\sigma = 0$) and negative $z$ direction ($\theta_\tau = \pi$), respectively. Note that under this conditions and $\theta_p=0$, the average velocity $\langle v_x\rangle$ barely develop oscillations (blue line), whiles $\langle v_y\rangle$ a clear oscillation occurs originated by the {\it transverse Zitterwebegung} with a period of $20\pi$ fs, which indicate that it is a valley-driven effect. On the other hand, when $\theta_p=\pi/2$ (green lines), the simultaneous presence of the pseudospin and valley induced {\it transverse Zitterwebegung} emerges in the transient behavior of $\langle v_x\rangle$, with periods of $2\pi$ fs and $20\pi$ fs, respectively. However, no oscillations are developed for the $\langle v_y\rangle$. The latter is due that the orientation of the momentum of the wave packet is parallel to the pseudospin and perpendicular to the valley polarization. In Fig. \ref{fig:FiguraVelocityZitterVariosAngulosMomentoV2}(c)-(d), we show the oscillatory contribution of the expected position of the wave packets, $ \langle \xi_x(t) \rangle $  and $ \langle \xi_y(t) \rangle $, respectively. Similarly, as it occurs in the average velocities, the average position follows the same dependence on the angle of the momentum.

\section{Concluding remarks} 

The emergent fermions in Kekul\'e-Y textured graphene are described by two distinct chiralities related to the sublattice and valley degrees of freedom. Using numerical methods to solve for the dynamics of Gaussian wave packets in the Schr\"odinger picture, it was shown that the broken chiral symmetry allows wave packets to propagate at two different Fermi velocities determined by their chiralities, while retaining their pseudo-relativistic behavior from pristine graphene.
The dynamics in the Heisenberg picture were also analyzed and it was found that the valley-momentum locking phenomenon introduces a second {\it Zitterwebegung} in addition to the well-known pseudospin {\it Zitterwebegung} found in pristine graphene. Similarly to the pseudospin {\it Zitterwebegung}, this new {\it Zitterwebegung}  can be interpreted as generated by the precession of the valley degree of freedom ($\tau$) around the momentum with a characteristic angular frequency $ \omega_\tau = \Delta_0\omega_\sigma $, where $ \Delta_0 \leq 1 $ is the Kekul\'e coupling constant and $ \omega_\sigma $ is the renormalized angular frequency of pseudospin {\it Zitterwebegung} in Kekul\'e-Y graphene. While the pseudospin {\it Zitterwebegung} is due to interference between fermionic and anti-fermionic energy states, this valley {\it Zitterwebegung} is due to the mixing of valley states with the same energy, akin to the {\it Zitterwebegung} type occuring in two-dimensional semiconductor systems with Rashba spin-orbit coupling.  The valley {\it Zitterwebegung} has been shown to share some of the properties as the pseudospin {\it Zitterwebegung} as its amplitude increases as the angle between the valley isospin and momentum increases, and it dampens to zero at larger times. However, the lower frequency of oscillation leads to a drastic decrease in the attenuation rate of valley {\it Zitterwebegung} compared to pseudospin {\it Zitterwebegung}. We showed that the appearance of these {\it Zitterbewegung} frequencies are closely related to the Berry connection matrix of the Kek-Y graphene Hamiltonian. The results opens up new routes that can be explored in Dirac materials with Kekulé distortions for the observation of the {\it Zitterbewegung} phenomenon.

\begin{acknowledgments}
A.S. and R.C-B. acknowledge useful discussions with Elias Andrade, Gerardo G. Naumis and Mahmoud M. Asmar. A.S, P.E.I. and R.C.-B. thank 20va Convocatoria Interna (UABC) 400/1/64/20. F.M. acknowledges the support of DGAPA-UNAM through the project PAPIIT No. IN113920.

\end{acknowledgments}

\bibliography{Xrefs.bib}
\end{document}